\documentclass{PoS}
\usepackage[utf8x]{inputenc}
\usepackage{amssymb}
\usepackage{amsmath}

\newcommand{\rt}{{\mathbf{r}}}

\newcommand{\bt}{{\mathbf{b}}}
\newcommand{\bti}{{\mathbf{b}_{i}}}
\newcommand{\gev}{\ \textrm{GeV}}
\newcommand{\der}{\mathrm{d}}
\newcommand{\xpom}{{x_\mathbb{P}}}
\newcommand{\A}{\mathcal{A}}

\title{Proton shape fluctuations and its relation to DIS}

\ShortTitle{Proton shape fluctuations}

\author{\speaker{Heikki Mäntysaari}\\
       Department of Physics, P.O. Box 35, 40014 University of Jyv\"askyl\"a, Finland\\
        E-mail: \email{heikki.mantysaari@jyu.fi}}

\abstract{We review the recent progress in extracting the proton fluctuating substructure by studying exclusive processes at HERA, and the applications of these developments in the interpretation of the LHC heavy ion data. The possibilities to extract the proton geometry directly from the LHC high-multiplicity  proton-nucleus and proton-proton collision data is also discussed. }

\FullConference{XXVI International Workshop on Deep-Inelastic Scattering and Related Subjects\\
		16-20 April 2018\\
		Kobe, Japan}

\begin{document}

\section{Introduction}

The success of hydrodynamical simulations in description of the vast amount of collective phenomena seen in the heavy ion collisions at RHIC and at the LHC has shown conclusively the production of a new state of matter, the quark gluon plasma (see e.g.~\cite{Schenke:2010rr,Gale:2012rq,Schenke:2012wb,Paatelainen:2013eea}). In these collisions, the produced locally thermal matter goes trough a hydrodynamical evolution, which transforms initial state spatial anisotropies into final state momentum space correlations. These can be quantified by decomposing the particle production spectra as a function of azimuthal angle into Fourier harmonics $v_n$. For a review of the relativistic hydrodynamics and the description of the heavy ion collisions, the reader is referred to Ref.~\cite{Gale:2013da}.

Studies of the proton-lead collisions have revealed a surprisingly similar collective effects, such as large elliptic and triangular flows ($v_2$ and $v_3$), than what was previously seen in heavy ion collisions. Traditionally results of this kind have been taken as a proof for the production of the collectively evolving quark gluon plasma~\cite{Chatrchyan:2013nka}. For a review of recent collectivity measurements, the reader is referred to Ref.~\cite{Dusling:2015gta}. This has raised a natural question whether it is possible the produce the (locally) thermal matter also in case of small collision systems.   

A crucial input for the hydrodynamical simulations is the description of the initial state. As shown in Ref.~\cite{Schenke:2014zha}, when the same so called IP-Glasma~\cite{Schenke:2012wb} framework that successfully describes the heavy ion data is applied to proton-nucleus collisions, too small elliptic flow is obtained compared with the CMS data~\cite{Schenke:2014zha}. However, these first simulations did not include any event-by-event fluctuations of the geometric shape of the proton that one can generally expect to be present in the quantum mechanical wave function of the proton. As pointed out in Ref.~\cite{Schenke:2014zha} the discrepancy between the simulations and the CMS data could be explained by having an initial state with large eccentricities. 

From the heavy ion phenomenology point of view, it is thus crucial to answer the fundamental question about the spatial distribution of quarks and gluons inside the proton. And more importantly, how do these distribution fluctuate on an event-by-event basis.  In the following, we discuss how this part of the proton wave function can be constrained by studying  exclusive vector meson production in deep inelastic scattering, or by analyzing the high multiplicity proton-nucleus and proton-proton data.

\section{Exclusive vector meson production as a probe of the proton structure}
\label{sec:diffraction}
In the Good-Walker picture~\cite{Good:1960ba}, diffractive scattering is described in terms of states that diagonalize the scattering matrix. 
Assuming that the center-of-mass energy is large and neglecting states that are suppressed by powers of the strong coupling $\alpha_s$, these states  are the ones where the virtual photon splits into a quark-antiquark dipole whose transverse separation is fixed, and the dipole probes the target in a configuration which does not change during the scattering process.

A convenient framework to describe inclusive and diffractive deep inelastic scattering processes at high energy is provided by the Color Glass Condensate framework~\cite{Gelis:2010nm}. Let us study the production of vector mesons at high energy, where the high energy factorization can be applied. In this case, the scattering amplitude for (in this case) $J/\Psi$ production can be factorized into three parts. First, the incoming virtual photon splits into a quark-antiquark dipole, which can be computed applying perturbative QED~\cite{Kovchegov:2012mbw} to get the photon wave function $\Psi$. Then, the dipole scatters elastically off the target hadron, and the amplitude for the dipole-target scattering is called the dipole amplitude $N(\rt,\bt,x)$, where $\rt$ is the transverse size of the dipole, $\bt$ is the impact parameter and $x$ the Bjorken-$x$ of DIS. Finally, the dipole forms a vector meson, the formation process being described in terms of the vector meson wave function $\Psi_V$.

As the scattering amplitude is proportional to the dipole amplitude $N$ which itself is approximately proportional to the gluon density inside the target, the exclusive vector meson production cross section is especially sensitive to the gluons being proportional to $N^2$. Similarly, in perturbative QCD calculations, the power of exclusive processes in probing the gluon densities is found by noticing that at leading order the fact that no net color is transferred requires one to exchange at least two gluons. For a more rigorous calculation, the reader is referred to Refs.~\cite{Ryskin:1992ui,Brodsky:1994kf}.

In addition to being especially sensitive to the parton densities, exclusive vector meson production has the advantage that it is possible to measure the transverse momentum transfer from the target. As the impact parameter is, by definition, the Fourier conjugate to the transverse momentum, it becomes possible to study the spatial distribution of the dipole-target scattering amplitude by measuring exclusive vector meson production.

The scattering amplitude for the exclusive vector meson production in the dipole picture reads~\cite{Kowalski:2006hc}
\begin{equation}
\label{eq:diff_amp}
 \A^{\gamma^* p \to V p}_{T,L}(\xpom,Q^2, \boldsymbol{\Delta}) = i\int \der^2 \rt \int \der^2 \bt \int \frac{\der z}{4\pi}  
  (\Psi^*\Psi_V)_{T,L}(Q^2, \rt,z) 
 e^{-i[\bt - (1-z)\rt]\cdot \boldsymbol{\Delta}}  2N(\rt,\bt,\xpom),
\end{equation}
where $\Psi^*\Psi_V$ denote the overlap between the virtual photon and the vector meson wave functions, and $z$ is longitudinal momentum fraction of the photon carried by the quark. We emphasise that as the total cross section can be obtained from the forward elastic scattering amplitude (corresponding to the zero transverse momentum transfer $\Delta$) by applying the optical theorem, the total cross section is only sensitive to the dipole amplitude $N$ integrated over the whole transverse area. In contrast, in order to calculate the scattering amplitude for the exclusive process \eqref{eq:diff_amp}, it is necessary to know the detailed impact parameter profile of the dipole-target scattering amplitude $N$, and thus the spatial density profile of the target.

Diffractive processes are divided into two categories. The events where the target proton remains intact are obtained by averaging the scattering amplitude~\eqref{eq:diff_amp} over the possible target configurations, and then taking the square~\cite{Kowalski:2006hc,Miettinen:1978jb}
\begin{equation}
\frac{\der \sigma^{\gamma^*p \to Vp}}{\der t} = \frac{1}{16\pi} \left | \left \langle   \A^{\gamma^* p \to V p}  \right \rangle \right|^2,
\end{equation}
which depends on $\langle N(\rt, \bt, b)\rangle$ and thus the average density profile of the proton. Here the average $\langle \rangle$ refers to averaging over possible configurations of the target, and $t \approx -\Delta^2$.

  \begin{figure}[tb]
    \centering
    \begin{minipage}{.48\textwidth}
        \centering
        \includegraphics[width=\textwidth]{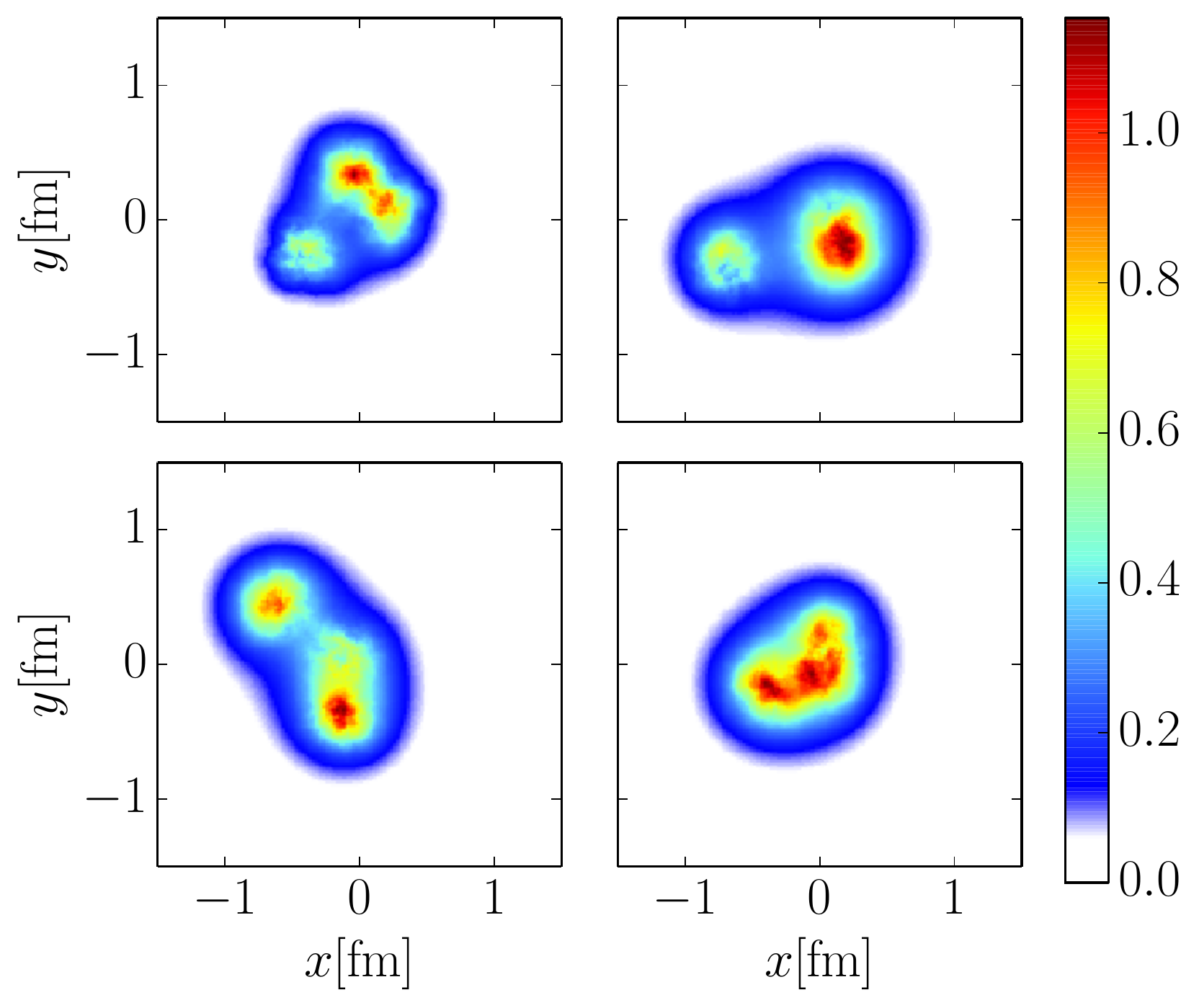} 
				\caption{Illustration of the event-by-event fluctuations of the proton density profile $1- \Re \mathrm{Tr} V(\bt)/N_c$.}
		\label{fig:ipglasma}
    \end{minipage}
    \quad
     \begin{minipage}{0.48\textwidth}
        \centering
      \includegraphics[width=\textwidth]{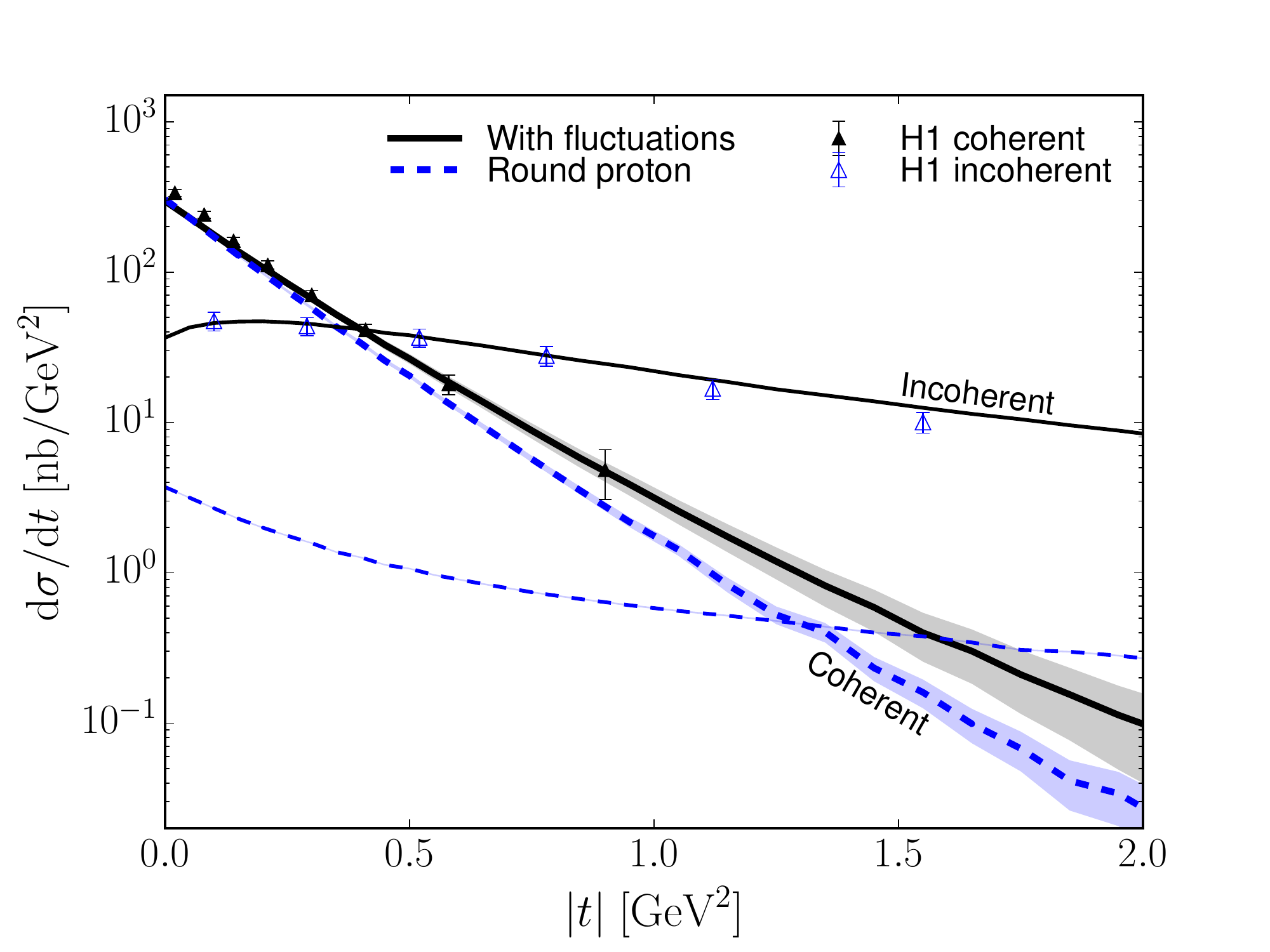} 
				\caption{Exclusive $J/\Psi$ photoproduction cross section at $W=75\gev$ compared with the H1 data~\cite{Alexa:2013xxa} with and without fluctuations of the proton shape and density.}
		\label{fig:jpsi_spectra}
    \end{minipage}

\end{figure}

On the other hand, if one calculates total diffractive cross section and subtracts the cross section for the coherent scattering, the resulting cross section corresponds to the case where the target dissociates. This we call incoherent scattering here (in case of both proton and nuclear targets), and the cross section becomes a variance
\begin{equation}
\frac{\der \sigma^{\gamma^*p \to Vp^*}}{\der t} = \frac{1}{16\pi} \left( \left \langle \left|   \A^{\gamma^* p \to V p}  \right|^2 \right \rangle -  \left | \left \langle   \A^{\gamma^* p \to V p}  \right \rangle \right|^2 \right).
\end{equation}
As a variance, the incoherent cross section measures the amount of fluctuations in the event-by-event distribution of the dipole amplitude, and thus in the density profile of the target~\cite{Miettinen:1978jb,Frankfurt:2008vi,Lappi:2010dd}. As $-\sqrt{t}$ is the Fourier conjugate to the impact parameter, at small $|t$| one is sensitive to fluctuations at long distance scales, and  shorter distances are probed at large $|t|$.

In Refs.~\cite{Mantysaari:2016ykx,Mantysaari:2016jaz} the fluctuating proton geometry is constrained by fitting the HERA exclusive $J/\Psi$ production data~\cite{Alexa:2013xxa}. In these works, the proton density profile is parametrized by two numbers: the width $B_{qc}$ of the distribution from which the positions for the constituent quark positions are sampled, and the width $B_q$ of the Gaussian distribution of gluon density around the sampled $3$ quarks. Additionally, the fluctuations of the overall density (or that of the saturation scale) are implemented by allowing the overall density of each of the \emph{hot spot} to fluctuate independently. The proton transverse density profile is written as
\begin{equation}
T_p(\bt) = \sum_{i=1}^3 T(\bt - \bti) \quad \text{with} \quad T_q(\bt) = \frac{\sigma_i}{3}\frac{1}{2\pi B_q} e^{-\bt^2/(2B_q)}.
\end{equation}
Here $\sigma_i$ corresponds to the density fluctuation at is sampled from a log-normal distribution, see Ref.~\cite{Mantysaari:2016jaz} for details.
 
When the proton density profile is known, in Refs.~\cite{Mantysaari:2016ykx,Mantysaari:2016jaz} the dipole-target scattering amplitude $N(\rt,\bt,x)$ is obtained by sampling the color charges from a distribution with expectation value being dictated by the density at the given point, and then solving the classical Yang-Mills equations. With optimal set of parameters the resulting photon shapes are illustrated in Fig.~\ref{fig:ipglasma}, where the real part of the trace of the Wilson line is shown\footnote{We note that the trace of a single Wilson line is not gauge invariant, but the point of Fig.~\ref{fig:ipglasma} is to illustrate the amount of geometrical fluctuations}. The description of the H1 data is show in Fig.~\ref{fig:jpsi_spectra} where, for comparison, also results obtained with only color charge fluctuations inside the round proton are shown. In both cases (with and without geometry fluctuations) the coherent cross section which measures the average density profile is well reproduced, but large geometry fluctuations are necessary in order to describe the incoherent cross section.

\section{Applications on proton-nucleus collisions}

The result of the analysis of Refs.~\cite{Mantysaari:2016ykx,Mantysaari:2016jaz} is the event-by-event distribution of the Wilson lines that describe the small-$x$ structure of the proton. These Wilson lines are an input for the simulations of the proton-nucleus collisions in the IP-Glasma framework (see Ref.~\cite{Schenke:2012wb}). In these simulations one first obtains the pre-thermal evolution by solving the classical Yang-Mills equations, and the obtained energy-momentum tensor is used to initialize the hydrodynamical simulations when the system is close to thermal equilibrium. When the system expands it eventually cools down and hadronizes, and the evolution in the hadronic phase is simulated by a microscopic afterburner UrQMD~\cite{Bleicher:1999xi}.

\begin{figure}[tb]
    \centering
    \begin{minipage}{.48\textwidth}
        \centering
        \includegraphics[width=\textwidth]{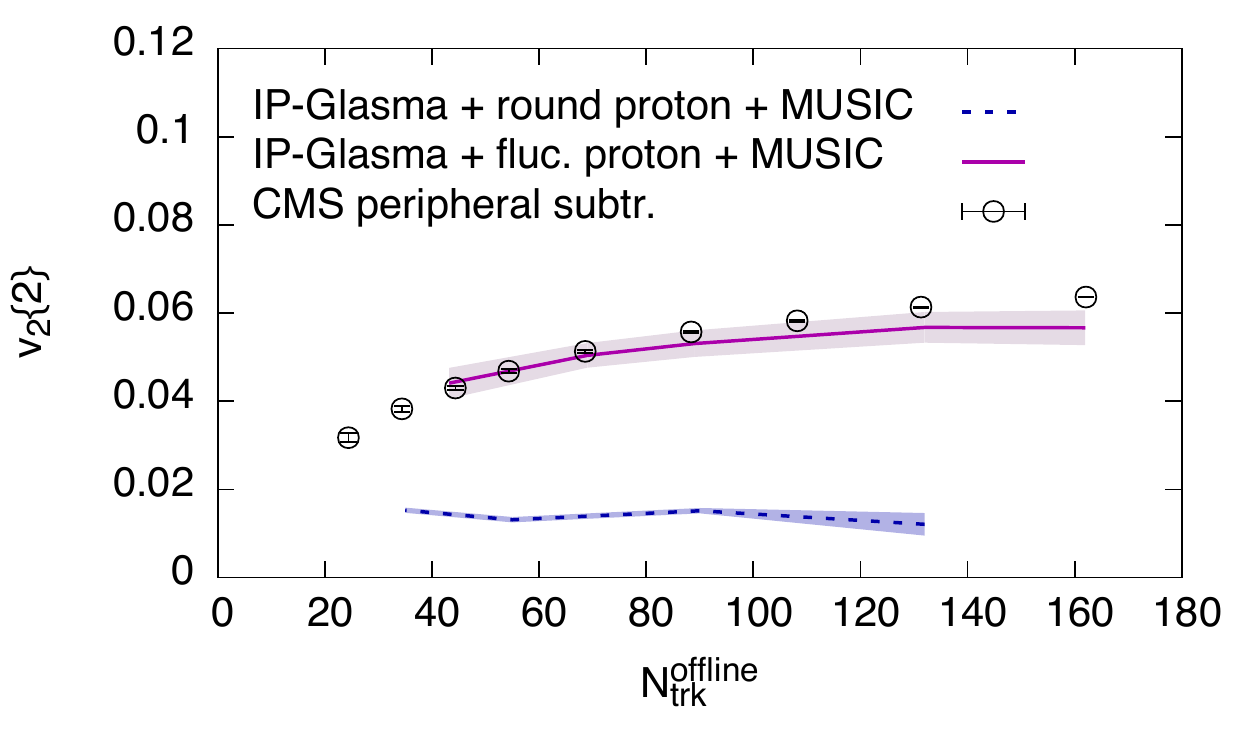} 
				\caption{Elliptic flow $v_2$ extracted using round protons (from \cite{Schenke:2014zha}) and protons with fluctuating substructure constrained by the HERA $J/\Psi$ data. Data is from the CMS collaboration~\cite{Chatrchyan:2013nka}}
		\label{fig:v2}
    \end{minipage} \quad
    \begin{minipage}{0.48\textwidth}
        \centering
      \includegraphics[width=\textwidth]{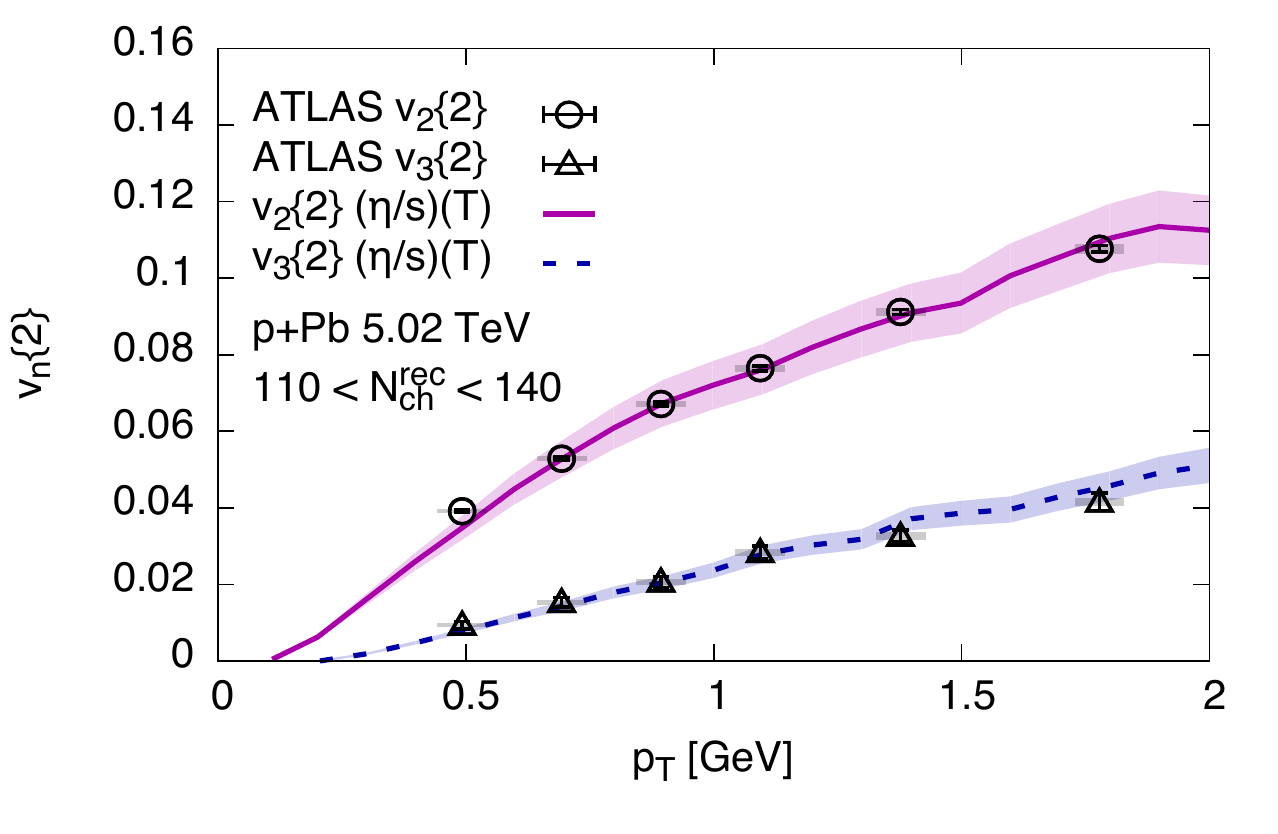} 
				\caption{Elliptic and triangular flow as a function of transverse momentum computed with the fluctuating proton and compared. with the ATLAS data~\cite{Aad:2014lta} (figure from Ref.~\cite{Mantysaari:2017cni}).}
		\label{fig:v2v3_pt}
    \end{minipage}
\end{figure}

The resulting elliptic flow as a function of event multiplicity (which is related to centrality in heavy ion collisions) reported in Ref.~\cite{Mantysaari:2017cni} is shown in Fig.~\ref{fig:v2}. For comparison, the result of Ref.~\cite{Schenke:2014zha} where the same framework is applied with round protons is shown. The geometry fluctuations constrained by the HERA data that generate large eccentricities in the initial state are found to be large enough the reproduce the CMS measurement~\cite{Chatrchyan:2013nka}. 

In addition to total $v_2$, there is more differential data from proton-nucleus collisions. As an example, in Fig.~\ref{fig:v2v3_pt} the transverse momentum dependence of the elliptic and triangular flow is shown and compared with the ATLAS data. Similarly, the description of the particle species dependence of the flow harmonics and the HBT radii were found to be described accurately in Ref.~\cite{Mantysaari:2017cni}.

Even though the hydro calculations are successful in describing the LHC data, it is important to realise that there are also purely initial state calculations that can potentially explain the observed flow-like behaviour. For example, in Refs.~\cite{Dusling:2017dqg,Mace:2018vwq} the gluon correlations at the initial state are found to explain the observed flow harmonics. For a review of the collective effects in proton-proton and proton-nucleus collisions, we refer the reader to~\cite{Dusling:2015gta}. Additionally, the applicability of the relativistic hydrodynamics on the evolution of small systems is not conclusively proven, see e.g.~\cite{Niemi:2014wta}.

\section{Accessing $x$ and system size dependence of the hadron shape fluctuations}

 \begin{figure}[tb]
    \centering
    \begin{minipage}{.48\textwidth}
        \centering
        \includegraphics[width=\textwidth]{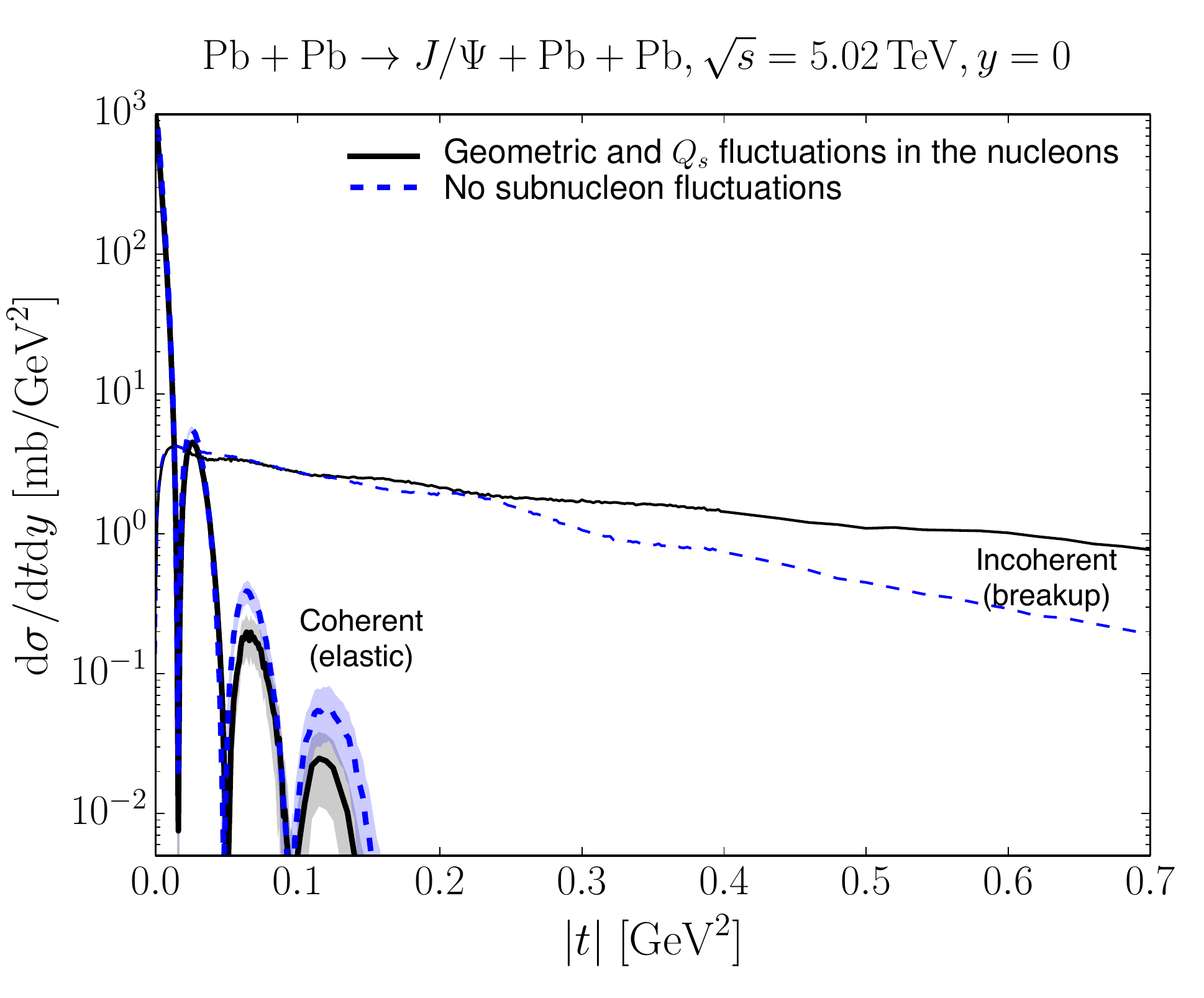} 
				\caption{$J/\Psi$ photoproduction spectra in ultraperipheral $Pb+Pb$ collisions at the LHC. Results with and without nucleon substructure fluctuations are shown. }
		\label{fig:upc_jpsi}
    \end{minipage} \quad
    \begin{minipage}{0.48\textwidth}
        \centering
      \includegraphics[width=0.9\textwidth]{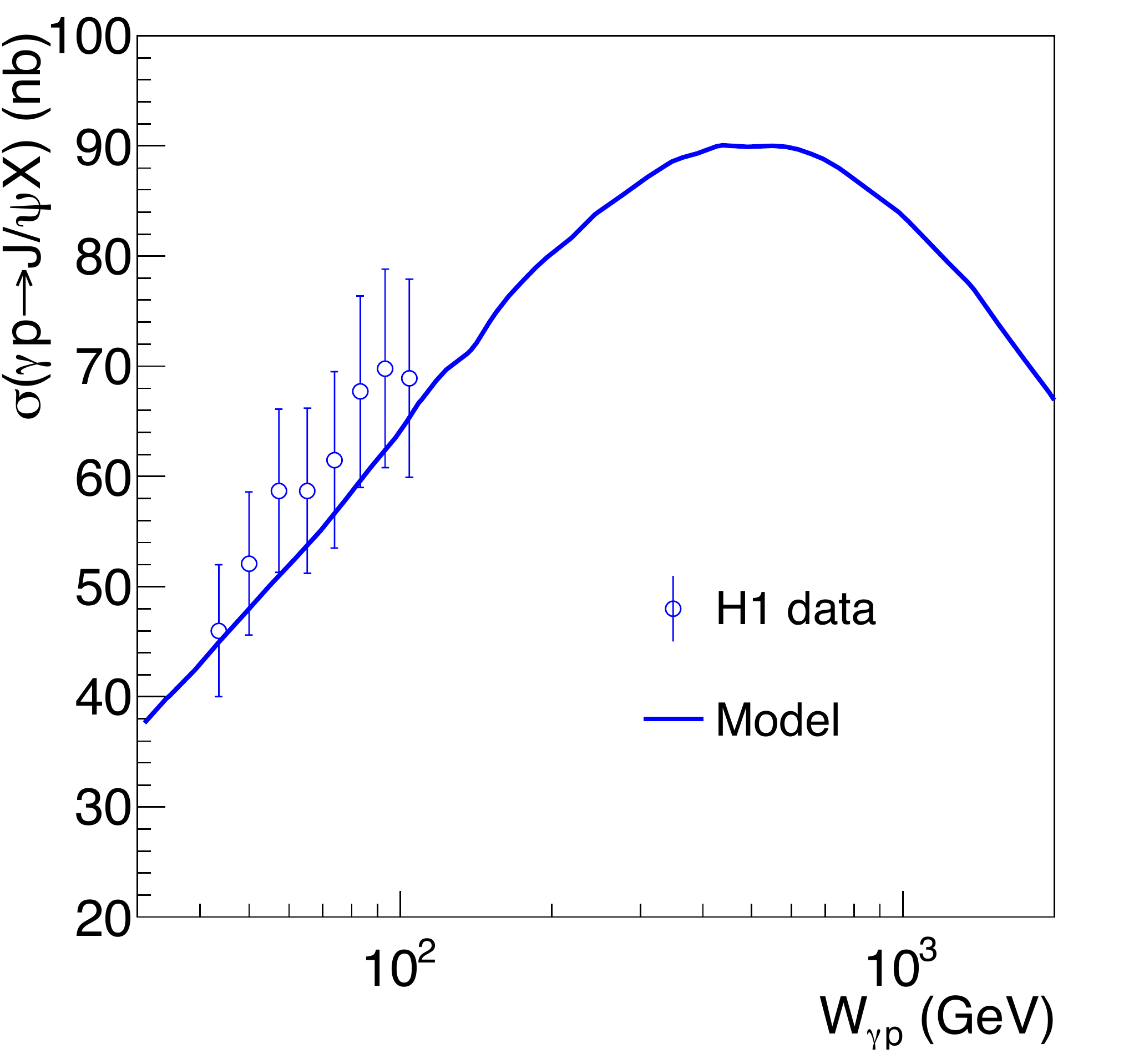} 
				\caption{Incoherent $J/\Psi$ production cross section as a function of center-of-mass energy in the model of Ref.~\cite{Cepila:2016uku} where the number of hot spots increasing with increasing energy.}
		\label{fig:incoh_wdep}
    \end{minipage}
\end{figure}

Recently ultraperipheral heavy ion collisions have provided us an access to study photoproduction processes at very small $x$, and with nuclear targets. In these events, the impact parameter is larger than the sum of the nuclear radii, which makes the dominant process to be the one where one of the nuclei acts as a source of almost real photons that interact with the other nucleus. As the photon flux is proportional to the electric charge squared, if one studies ultraperipheral proton-nucleus collisions at the LHC one is effectively studying photon-proton scattering at potentially very small $x$ (and at $Q^2 \approx 0$). For a review, see Ref.~\cite{Bertulani:2005ru}.

As discussed in Sec.~\ref{sec:diffraction}, the incoherent vector meson production probes fluctuations in short and long distance scale depending on $t$. This is demonstrated in Ref.~\cite{Mantysaari:2017dwh}, where the photoproduction of $J/\Psi$ off nuclei was studied with nucleon shape fluctuations constrained by the HERA data. The $J/\Psi$ photoproduction spectra in ultraperipheral Pb$+$Pb collisions is shown in Fig.~\ref{fig:upc_jpsi}, where the coherent and incoherent cross sections are computed with and without nucleon geometry fluctuations. In both cases the coherent cross sections are identical, as the average nuclear geometry is not affected by the inclusion of the nucleon shape fluctuations. Similarly, at small $|t|$ where fluctuations at long distance scales (fluctuating positions of the nucleons from the Woods-Saxon distribution) are probed the obtained incoherent spectra and similar. However, at large momentum transfer (which is well within the LHC kinematical reach) the incoherent cross sections start to be vastly different, and the subnucleon scale fluctuations contribute significantly.

Measuring $J/\Psi$ production as a function of rapidity in ultraperipheral proton-nucleus collisions makes it possible to study the $x$ dependence of the proton structure. In these events the nucelus dominantly acts as a photon source as discussed above. The Bjorken-$x$ probed in the production of a vector meson with mass $M_V$ then becomes $\xpom = M_V e^{-y}/\sqrt{s}$, and there is no ambiguity in determining which object acted as a photon source, and the kinematics can be uniquely constructed.

In the DIS 2018 meeting, the ALICE collaboration reported new measurements of the coherent and incoherent $J/\Psi$ production at different photon-proton center of mass energies~\cite{ContrerasDIS2018}. These measurements were in line with previous results from run 1~\cite{TheALICE:2014dwa}. In these datasets, the incoherent cross section becomes unobservable when one reaches highest center of mass energies. The data analysis is not final and results are not reported at the cross section level, but this hints towards the approach of the black disk ilimit already in the LHC kinematics, where the highest energy bin corresponds to photon-proton center of mass energy $W \sim 800\gev$ as shown in Fig.~\ref{fig:incoh_wdep}. This is due to the general feature that when the black disk limit is reached and the proton has becomes fully saturated, there can not be any fluctuations in the scattering amplitude and thus the incoherent cross section vanishes.

A phenomenological model to explain this disappearance from the dipole picture was presented in the conference by J. G. Contreras~\cite{Cepila:2018zky}. In this work, which is based on their initial article~\cite{Cepila:2016uku} the proton structure evolution is modelled by implementing a proton substructure which consist of an energy-dependent number of hot spots. When the model parameters are fixed to the data at HERA kinematics, the authors predict an incoherent cross section which starts to decrease around the maximum LHC energies.

A second approach to describe the energy dependence is to solve small-$x$ evolution equations (BK or JIMWLK)  with impact parameter dependence similarly as in Refs.~\cite{Berger:2012wx,Schlichting:2014ipa}. These calculations are able to  explain also the growth of the proton with decreasing $x$ observed at the HERA data, and the fact that the coherent cross section becomes dominant at high energies, but are challenging due to the presence of long distance scale (compared to the size of the proton and $\Lambda_\text{QCD}^{-1}$) that must be modeled by an effective description of the confinement scale physics. First results based on this approach were presented in the conference~\cite{Mantysaari:2018zdd}.

 \begin{figure}[tb]
    \centering
    \begin{minipage}{0.48\textwidth}
        \centering
      \includegraphics[width=0.8\textwidth]{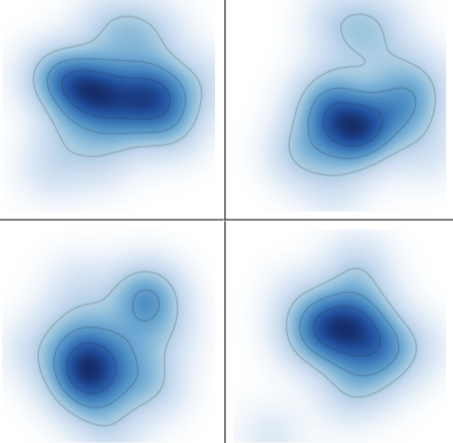} 
				\caption{Extracted proton density profiles by a Bayesian analysis. Figure from Ref.~\cite{MorelandQM2017}.}
		\label{fig:bayesian_protons}
    \end{minipage} \quad
    \begin{minipage}{0.48\textwidth}
        \centering
      \includegraphics[width=1.1\textwidth]{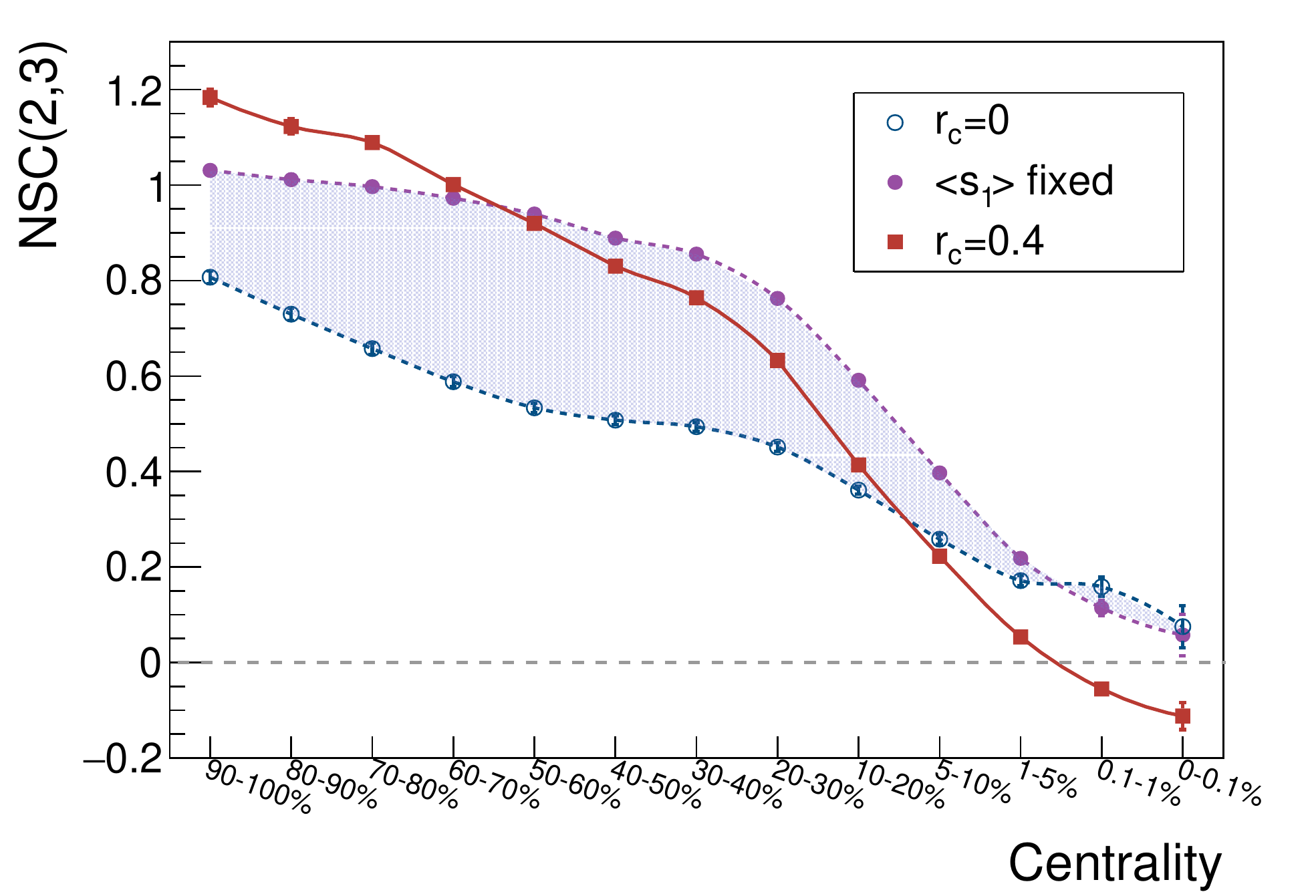} 
				\caption{Correlator between the flow harmonics $v_2$ and $v_3$ from~\cite{Albacete:2017ajt}. The red $r_c=0.4$ line corresponds to the case where the hot spots have short range repulsive correlation, which turns the correlation function negative at most central collisions.  }
		\label{fig:v23_correlation}
    \end{minipage}
\end{figure}

\section{Extracting shape fluctuations from proton-nucleus or proton-proton collisions}

An another approach to determine the proton substructure developed recently is based on Bayesian statistics. In the analysis of Ref.~\cite{Moreland:2017kdx}, the authors perform multiple hydrodynamical simulations of the proton-nucleus collisions with different proton thickness functions on an event-by-event basis. The Bayesian statistics makes it then possible to extract posterior distributions for the proton thickness function. Examples of the extracted proton profiles are shown in Fig.~\ref{fig:bayesian_protons}, the results being qualitatively similar to the ones obtained by studying $J/\Psi$ production at HERA and shown in Fig.~\ref{fig:ipglasma}.

After DIS2018, the authors have also presented up-to-dated analysis~\cite{MorelandQM2018}. These analyses prefer initial entropy deposition to be similar as in case of IP-Glasma framewok used in Ref.~\cite{Mantysaari:2017cni}. The actual number of hot spots can not be constrained by the Bayesian analysis with current data, but tjhe case of a singe constituent (meaning no substructure) is strongly disfavored. Quatitatively the largest difference to the proton substructure constrained by the diffractive data in Ref~\cite{Mantysaari:2016jaz} is the proton transverse area for which the Bayesian analysis prefers much larger values than the vector meson production data. However, it is not a priori clear that the proton transverse area which controls the initial entropy deposition in proton-nucleus collisions is exactly the same as the transverse distribution of the small-$x$ gluons probed in $J/\Psi$ production.

In the calculations presented above there are no correlations between the hot spots positions. One approach to go beyond this approximation is presented in Ref.~\cite{Albacete:2017ajt}, where the idea is to look for correlator between the $v_2$ and $v_3$ harmonic flow coefficients in high multiplicity proton-proton collisions.  As observed by CMS~\cite{Sirunyan:2017uyl}, this correlator becomes negative at very high multiplicities. In order to explain this behaviour is is found that in the proton which consist of fluctuating hot spots, there has to be a repulsive short range correlation between them (see also Ref.~\cite{Traini:2018hxd} for a discussion about the quark position correlations). The obtained $v_2,v_3$ correlator is shown in Fig.~\ref{fig:v23_correlation}. The same framework has also been used to explain the so called hollowness effect in elastic proton-proton collisions in Ref.~\cite{Albacete:2016pmp}.

\section{Conclusions}

Based on HERA and LHC data, it has become possible to take multi dimensional snapshots of the proton substructure on an event by event basis. The proton substructure constrained by the HERA data has significant implications on the interpretation of the LHC data on high multiplicity proton-nucleus and proton-proton collisions. In particular, they suggest that the observed collective effects in the proton-lead and possibly even in proton-proton collisions might be of hydrodynmaical origin, and the proton geometry required to explain the LHC data in hydro simulations is compatible with the  proton shape fluctuations extracted from exclusive HERA data.

Even though there are other purely initial state explanations of the high-multiplicity proton-nucleus data (e.g.~\cite{Dusling:2017dqg}), the HERA proton dissociative $J/\Psi$ production cross section generally suggest that there has to be large event-by-event fluctuations in the dipole-target scattering amplitude. The power of exclusive processes is that, by definition,  impact parameter is the Fourier conjugate to the total transverse momentum transfer which is measurable in these events, providing direct access to target geometry. As discussed in detail in Refs.~\cite{Mantysaari:2016ykx,Mantysaari:2016jaz,Mantysaari:2017dwh}, this data is explained in a picture with large geometric fluctuations, and  overall density fluctuations on top of that. 

The possibility to measure diffractive vector meson production off a proton at very high energies in  ultraperipheral proton-lead collisions has also provided us strong hints of the Bjorken-$x$ evolution of the fluctuating proton structure. In particular, the recent LHC data suggests that the fluctuations get reduced significantly which hints towards the black disk limit of the proton being approached in the current LHC kinematics. If confirmed, this would provide  support to the presence of gluon saturation in the current collider energies.

\section*{Acknowledgments}
 H. M. is supported by European Research Council, Grant ERC-2015-CoG-681707.
 
\bibliographystyle{JHEP}
\bibliography{../../refs}
\end{document}